\documentstyle[preprint,aps]{revtex}
\begin{document}
\preprint{cond-mat/9501007}

\draft
\title{Stability of Chiral
Luttinger Liquids and Abelian Quantum Hall States}
\author{F. D. M. Haldane}
\address{Department of Physics, Princeton University, Princeton, New Jersey
08544}
\date{November 23, 1994; Revised December 31, 1994}
\maketitle
\begin{abstract}
A criterion is given for  topological stability of Abelian quantum Hall
states, and of Luttinger liquids at the boundaries between
such states; this suggests a selection rule on
states in the quantum Hall hierarchy theory.
The linear response of  Luttinger
liquids to electromagnetic fields is described:
the Hall conductance is quantized, irrespective
of whether edge modes propagate in different directions.
\end{abstract}


\narrowtext
A key feature of the quantum Hall effect (QHE) is its stability
with respect to impurity potentials.
A region of incompressible
quantum Hall fluid described by a simple
Laughlin state\cite{laughlin}
has a current-carrying boundary where gapless excitations
can be made.  The edge is gapless
because the current is carried
by a single {\it chiral} ``Luttinger liquid'' mode\cite{wen}:
stability against impurity perturbations may be associated with
the {\it maximal chirality} of the edge. In contrast
to the case of a non-chiral Luttinger liquid\cite{haldane},
there are no states
of opposite chirality, so there is no backscattering (which can
drive a ``mass-generating''
instability in which a pair of gapless
modes with opposite chirality annihilate each other).
In the integer
QHE, there are a number of edge modes, but
all have the same chirality, so the edge is stable.
However, absolute stability (in the sense that there is
{\it no} backscattering process) can occur even if the
edge is {\it not} maximally-chiral; in this Letter, I
characterize  the criterion for this more precisely.
The property of edge stability also suggests a related
stability property for ``bulk'' QHE states.

Recently, it has been suggested by
Kane, Fisher and Polchinski\cite{kfp} (KFP)
that when a ``clean'' Hall edge is not maximally chiral,
the quantization of the Hall conductance generally breaks down
(KFP also claim that,
in certain - but not all - circumstances, ``randomness at the edge''
drives a novel non-mass-generating instability that
preempts this).   The coupling of non-maximally-chiral Luttinger liquids
to electromagnetic fields is examined here, and in contrast to
\cite{kfp}, {\it no} breakdown
of quantization of the Hall conductance of
``clean'' edges is found.

The essential ingredient of the effective low-energy
theory of a quantum Hall state is a charged Abelian Chern-Simons
(CS) gauge field\cite{girvin,kiv,read}
that couples both to the electromagnetic field
and any other neutral degrees of freedom of the
theory.
In a wide class of possible QHE states (Abelian quantum Hall
states), such as those of the hierarchy theory\cite{hald,halp},
the neutral degrees of freedom are also Abelian CS
fields\cite{read2,blokwen,frozee,wenzee}, and
the Lagrangian density is given by
\begin{equation}
2\pi{\cal L} = \epsilon^{\lambda\mu\nu} \left (\mbox{$\case{1}{2}$}
\hbar ({\bf a}_{\lambda},{\bf K}\partial_{\mu}
{\bf a}_{\nu}) + eA_{\lambda} ({\bf q},\partial_{\mu} {\bf a}_{\nu}) \right ),
\label{cslagrangian}
\end{equation}
where ${\bf a}_{\mu}$ is an $n$-component vector of Abelian CS gauge fields,
$A_{\mu}$ is the electromagnetic gauge field, ${\bf K}$ is a nonsingular
{\it integer} coupling matrix (see, {\it e.g.,} \cite{wenzee}),
${\bf q}$ is an {\it integer} vector; $({\bf a},{\bf b})$ is the inner product.
This is a {\it topological field theory}
(the Hamiltonian vanishes identically).

We may take ${\bf q}$ to be a multiple $q_0$ of a {\it primitive}
integer vector (one where the components have no common factor).
In a basis that separates off the neutral CS fields,
\begin{equation}
{\bf K} = \left (
\begin{array}{cc}
k_0 & {\bf k}^T \\
{\bf k} & {\bf K}_0
\end{array}
\right )
, \quad {\bf q} =
\left (
\begin{array}{c}
q_0 \\ {\bf 0}
\end{array}
\right ) ,
\end{equation}
where ${\bf K}_0$ is a integer matrix of dimension $(n-1)$. Then
$\sigma^H$ = $ \nu e^2 /h $, with \cite{read2,wenzee}
\begin{equation}
\nu =  ({\bf q},{\bf K}^{-1}{\bf q})
=  { q_0^2 \det K_0 \over \det K } .
\end{equation}
If the effective theory represents microscopic physics where electrons
are the only mobile particles, $k_0 + q_0$ must be even, and
${\bf K}_0$ must be an {\it even integer matrix},
which means that all its
diagonal elements are even.
The structure of the theory is invariant under an equivalence
transformation ${\bf K}$ $\rightarrow$ ${\bf WKW}^T$, ${\bf q} $
$\rightarrow $ ${\bf Wq}$, where ${\bf W}$ is an integer
matrix with $|\det W| $ = 1 \cite{read2}.

It will be useful to introduce the notion of the {\it primitive
form} of such a  theory.
If $\{ {\bf K},{\bf q}\}$ is not
in primitive form,
$\det K $ has a factor $p^2 > 1$, and  the theory
is related to a primitive form
$\{\tilde{\bf K},\tilde{\bf q}\}$ by an integer matrix
${\bf W}$ with $\det W$ = $p$, so that
$\bf K$ = ${\bf W}\tilde{\bf K}{\bf W}^T$, ${\bf q}$ =
${\bf W}\tilde{\bf q}$, and
$q_0$ = $\tilde{q}_0$.
The total  flux
${\bf \Phi}$ of the CS fields
is $\int d^2r\, \vec{\nabla} \times \vec{\bf a}$;
the effective theory allows vortex defects of the CS fields carrying flux
$2\pi{\bf n}$,  with ${\bf n}$  an integer vector, which
have core energies determined by a Hamiltonian not specified in
(\ref{cslagrangian}).
The electric
charge $Q$ is $e({\bf q},{\bf K}^{-1}{\bf n})$, and
$|\det K|$ classes of vortices
may be characterized by the value of
${\bf K}^{-1}{\bf n}$ modulo  integer vectors.
Vortices in the class where this vanishes
have conventional statistics and  integral $Q$ in
units $q_0e$.   The theory is in primitive form
if this is the {\it only} class with this property.
The primitive form of a theory is equivalent to a non-primitive
form with a restriction on the allowed classes of
vortices.

I  now propose a definition of {\it topological
stability} (T-stability): {\it an Abelian quantum Hall theory
is T-stable if and only if ${\bf K}_0$ does not represent zero}.
This means that there is no non-trivial solution of the
equation
$({\bf m},{\bf K}_0 {\bf m})$ = 0, or, equivalently,
of $({\bf m},{\bf Km})$ = $({\bf q},{\bf m})$ = 0.  The physical idea
motivating this is that if a theory is {\it not } T-stable,
a pair of CS fields with opposite chirality decouple from the others
and do not contribute to the QHE;
backscattering between  edge modes derived from these fields can
then drive a mass-generating  instability.
A theory that is not T-stable
has a primitive form
\begin{equation}
{\bf K} = \left (
\begin{array} {ccc}
0       & 1       & {\bf 0}^T \\
1       & k       & {\bf 0}^T \\
{\bf 0} & {\bf 0} & {\bf K}'
\end{array}
\right ),
\quad
{\bf q} = \left (
\begin{array} {c}
0 \\  q_0r\\ {\bf q}'
\end{array}
\right ) ,
\label{reduced}
\end{equation}
where $k$ = $q_0r \bmod 2$.
This is because ${\bf K}$ can always be written as a
symmetric tridiagonal matrix
with positive off-diagonal elements,
with $K_{11}$ = 0 if it represents zero\cite{wat},
and here  $q_1$ = 0 also.
If the theory is in primitive form, $K_{12}$ = 1.
A multiple of the first basis vector $|1\rangle$
(with which ${\bf K}$ represents zero)
can be added to $|3\rangle$
so $K_{23}$ vanishes, and to $|2\rangle$ to reduce $K_{22}$ to 0 or 1.
Then $\det K'$ = $-\det K$,
$\nu $ = $({\bf q}',{\bf K}'^{-1}{\bf q}')$, the signatures
$\sigma( K)$ and $\sigma (K')$ are the same, and $(q_0)'$
= $q_0r'$ ($r$ and $r'$ integral).

At the edge of a region of incompressible
Hall fluid, or more generally
at the boundary between  Hall fluids with
different $\sigma^H$, there
must be a charge-non-conserving edge current
with an
anomaly in its continuity equation:
\begin{equation}
\partial_t \rho (x) + \partial_x j(x) = \sigma^H  E_x,
\label{ce}
\end{equation}
where $x$ is the coordinate along the edge, and $\sigma^H$ =
$(e^2/ h)\Delta \nu$ is now the
{\it change} in Hall constant across the edge.  This is
compensates the jump in the Hall current normal to the edge
($E_x$ is continuous, but the Hall constant is not).
The edge  between  Abelian Hall fluids is
a ``Luttinger liquid''\cite{wen}, also
characterized by a $\{{\bf K},{\bf q}\}$ pair,
which is that of the bulk CS theory if the
edge is between a  QHE state and a non-conducting region.

The Luttinger liquid
is described in terms of $n$ fields $\varphi_i(x)$,
with equal-time
commutation relations
\begin{equation}
[\varphi_i(x),\varphi_j(x')] = i\pi \left (K_{ij}{\rm sgn}(x-x') + L_{ij}
\right )
\end{equation}
where $L_{ij}$ =
${\rm sgn}(i-j)(K_{ij} + q_iq_j)$ is a Klein factor.
Let
$\varphi_{\bf m}$ be $\sum_i m_i\varphi_i$, where
${\bf m}$ is integral.
A set of integrally-charged, fermionic or bosonic {\it local fields}
are given by
\begin{equation}
\Psi_{\bf m}(x) = e^{-i \varphi_{\bf m}}, \quad
[Q,\Psi_{\bf m}] = eq({\bf m})\Psi_{\bf m},
\end{equation}
with $q({\bf m})$ = $({\bf q},{\bf m})$.
These local fields
are also characterized by the integer  quadratic form
$K({\bf m})$  = $({\bf m},{\bf Km} )$ where
$(-1)^{K({\bf m})}$ = $(-1)^{q({\bf m})}$, and are fermionic
or bosonic depending on whether
$K({\bf m})$ is odd or even:
$\Psi_{\bf m}(x) \Psi_{\bf m'}(x')$
= $(-1)^{q({\bf m})q({\bf m}')}\Psi_{\bf m'}(x') \Psi_{\bf m}(x)$
for $x \ne x'$ (the Klein factor ensures this when ${\bf m} \ne {\bf m}'$).

The gauge-invariant electric charge density is
\begin{equation}
\rho  = { e \over 2 \pi } \sum_{ij} q_i K_{ij}^{-1}D_x \varphi_j ,
\end{equation}
where $D_x\varphi_i$ is the covariant derivative
$(\partial_x\varphi_i - q_ieA_x/\hbar)$,
and $A_x(x)$ is the component of the electromagnetic
vector potential parallel to the edge.  The
charge density operator has the anomalous
commutation
relation  $[\rho(x),\rho(x')]$ = $i\hbar \sigma^H\delta' (x-x')$, which
gives rise to the anomaly in the continuity equation.
The effective Hamiltonian is
$H$ = $H_{LL}$ + $\int dx \, \phi \rho $, where $\phi(x)$ is the electrostatic
potential, and
$H_{LL}$ = $H_0$ + $H_1$ is the
harmonic effective  Luttinger-liquid Hamiltonian:
\begin{eqnarray}
H_0 &=& {1 \over 2\pi} \int dx \, \epsilon_i(x) D_x\varphi_i(x) ,\\
H_1 &=& { 1\over 4\pi } \int dx \int dx'\, {\cal V}_{ij}(x,x') D_x\varphi_i(x)
D_{x'}\varphi_j(x') .
\end{eqnarray}
It will be convenient to consider the case where $H_{LL}$ is translationally
invariant, and long-range forces are screened, so that
$\epsilon_i(x)$ = $\hbar \omega_i$, constant, and ${\cal V}_{ij}(x,x')$ =
$\hbar V_{ij}\delta(x-x')$,
where $V_{ij}$ is a positive definite matrix with the dimensions of velocity.

The gauge-invariant edge current is
\begin{equation}
j = -{\delta H_{LL} \over \delta A_x} =
{ e \over 2 \pi } \sum_{ij} q_i \left ( \omega_i\delta_{ij}
+ V_{ij}D_x\varphi_j
\right ),
\label{current}
\end{equation}
and is not conserved if $\sigma^H \ne 0$.
In that case,
it is also possible to define a conserved but non-gauge-invariant
edge charge density and current by
$\rho^{tot}$ = $ \rho + \sigma^H A_x $, $j^{tot}$ =
$j + \sigma^H\phi$;
only  the {\it full} theory combining the
chiral Luttinger liquid with the CS fields of the bulk Hall fluids is both
gauge-invariant {\it and} current-conserving.

When diagonalized, $H_{LL}$ describes edge modes propagating with velocities
$v_{\lambda}$ obtained by solution of the generalized real
symmetric eigenvalue problem
\begin{equation}
{\bf V u}_{\lambda} = v_{\lambda} {\bf K}^{-1}{\bf u}_{\lambda} ,
\end{equation}
which has real eigenvalues and linearly-independent eigenvectors
when ${\bf V}$ is positive-definite.  It is convenient to define
the orthogonal matrix ${\bf O}$ by
\begin{equation}
{\bf V}^{1/2} {\bf K} {\bf V}^{1/2} = {\bf O} {\bf v}_d {\bf O}^T ,
\label{keqn}
\end{equation}
where ${\bf v}_d$ is the diagonal matrix ${\rm diag}(v_1,\ldots , v_n)$;
note that its signature is given by $\sigma (v_d)$ = $\sigma (K)$.

Since the effective Hamiltonian
is harmonic, it is straightforward to compute the
linear response of the
gauge-invariant edge charge density and current
to changes $\delta \phi (x,t)$, $\delta A_x(x,t)$
in the electromagnetic potentials (Kubo formula).
In the thermodynamic limit, $\delta \langle j(x,t) \rangle$
is  given by
\begin{equation}
 \int_{-\infty}^tdt'\int_{-\infty}^{\infty}
dx' \, \sigma^{xx}(x-x',t-t') \delta E_x(x',t'),
\label{kubo}
\end{equation}
and $\delta \langle \rho  \rangle$ is
given by  a similar expression  with a kernel
$\sigma^H(x,t)$, where (independent of temperature)
\begin{eqnarray}
\sigma^{xx}(x,t) &=&
{e^2\over h} \sum_{\lambda} (a_{\lambda})^2 \delta (x-v_{\lambda}t ),
\nonumber \\
\sigma^H(x,t) &=&
{e^2\over h} \sum_{\lambda} (a_{\lambda})^2 v_{\lambda}^{-1}
\delta (x-v_{\lambda}t ),
\end{eqnarray}
with $a_{\lambda}$ = $\sum_i q_i (V^{1/2}O)_{i\lambda}$.
Then $\sigma^H(x,0) $ = $\sigma^H\delta (x)$, and $\sigma^{xx}(x,0)$ =
$\gamma_D\delta(x)$ where $\gamma_D$ = $(e^2/h)({\bf q},{\bf Vq})$
is the Drude weight, which characterizes the commutation relation
$[\rho(x),j(x')]$ = $ i\hbar \gamma_D\delta'(x-x')$.

Now consider
\begin{equation}
\int_0^{\infty}dt \, \sigma^{xx}(x,t) =
G + \mbox{$\case{1}{2}$}\sigma^H{\rm sgn}(x),
\end{equation}
where $G$ = $\case{1}{2}(e^2/h) ({\bf q},{\bf \Lambda}^{-1} {\bf q} )$,
and
\begin{equation}
{\bf V}^{1/2}{\bf \Lambda }{\bf V}^{1/2} = {\bf O} |{\bf v}_d|{\bf O}^T ,
\label{leqn}
\end{equation}
with $|{\bf v}_d|$ $\equiv$ ${\rm diag}(|v_1|,\ldots ,|v_n|)$.
When $\sigma^H$ = 0, $G$ is the {\it conductance}\cite{apelrice}:
in steady state,  a uniform current $\delta \langle j \rangle $
= $G\int dx\, \delta E_x(x)$ flows in response to an applied potential drop
along the system.   This is {\it not} so when $\sigma_H \ne 0 $,
{\it because of the anomaly} (\ref{ce}) {\it which does not permit
a uniform steady-state current unless $\delta E_x$ vanishes}:
Hall currents flow onto
the boundary and build up a charge distribution that cancels the
externally-applied electric field, so that
the {\it total}
field $\delta E_x$ (which is what enters in the Kubo formula (\ref{kubo}))
vanishes in steady state.   Thus
the  Hall edge adjusts
itself to lie along an equipotential, and
the steady-state
gauge-invariant current $\delta \langle j \rangle $ vanishes.
The conserved but non-gauge-invariant current is then
constant, and given
by $\delta \langle j ^{tot} \rangle $ = $ \sigma^H \delta \phi $,
as  expected.

As a formal quantity, $G$
satisfies the inequality $G \ge \case{1}{2}|\sigma_H|$,
which is an equality when $a_{\lambda}$ is only nonzero
for modes $\lambda$ of one chirality.  In general, $G$
depends on ${\bf V}$ and is non-universal; however, the following
analysis shows that $G$ does {\it not} affect
the Hall conductance.

The expression (\ref{current}) is not valid at points $x_i$  where
currents $\delta i_i$ flow onto the Hall edge through tunneling contacts
(in the absence of a microscopic description of the contacts,
a Kubo formula for $\delta \langle j (x,t) \rangle $
can only be used when $x$ $\ne$ $x_i$).
In steady state,  the contact points $x_i$
are the only places that give  non-zero
contributions $-\delta V_i$ to $\int dx \, \delta E_x$.  The
Kubo formula gives $\delta i_i$ = $\delta j(x_i^+)$
$-$ $\delta j(x_i^-)$ = $\sigma^H \delta V_i $ {\it independent of $G$}.
Note that $\delta V_i$
is the potential jump  across the  contact {\it along the Hall edge},
{\it not} the potential difference between the edge and the reservoir
from
which the tunneling current is drawn, and {\it does not allow
the terminal  conductance of the contact to be deduced.}
If $x_1 < x_2$
$< x_3 < x_4$, and a measured current $\delta i $
flows onto the edge at $x_1$, and off it
at $x_3$, a potentiometer (drawing no current)
connected between $x_2$ and $x_4$
will measure a {\it quantized}  Hall conductance
$\delta i /  \delta V_{24} $ = $\sigma^H $.   This analysis
assumes that
the separation between
contacts (and perhaps their size)
is large compared to the short-distance cutoff scale above
which the effective Luttinger liquid theory is valid.

The {\it scaling dimension} of the field $\Psi_{\bf m}$ is defined
by the multiplicative renormalization when it is normal-ordered
with respect to the ground state of $H_{LL}$. If periodic
boundary conditions on a length $L$ are imposed,
$:\Psi_{\bf m}:$ = $(L/\xi)^{\Delta ({\bf m})}\Psi_{\bf m}$, where
$\Delta ({\bf m})$ = $\case{1}{2} ({\bf m},{\bf \Lambda m} )$
($\xi$ is a short-distance cutoff). From (\ref{keqn}) and (\ref{leqn}),
$\Delta ({\bf m}) \ge \case{1}{2} |K({\bf m})|$.

Now consider stability of the Luttinger liquid with respect to perturbations
of the form
\begin{equation}
H' = \int dx \,
\left ( t(x)\Psi_{\bf m}(x) + t^*(x)\Psi_{-\bf m}(x) \right ).
\end{equation}
Three cases can be considered; (a) {\it clean} case:
$t(x)$ =  $|t(x)|\exp i \alpha (x) $
where $|t(x)|$ and $\partial_x \alpha(x)$ are constant, (b)
{\it random } case: they vary randomly, with
$\langle |t(x)|^2 \rangle $ finite, and (c) {\it local} case:
$|t(x)|$ vanishes except near a point $x_0$.

Clearly $\Psi_{\bf m}$ must be bosonic and charge-conserving,
with $q({\bf m}) $ = 0 and even $K({\bf m})$.   In the interaction
picture,
the vacuum amplitude
\begin{equation}
U(L,T) = \langle 0 | T_t \exp -i\hbar^{-1}\int_0^Tdt\, H'(t) |0 \rangle
\end{equation}
can be expanded
in powers of $t(x)$:
\begin{equation}
 \sum_{n=0}^{\infty} {(-i)^n \over n!} \prod_{i=1}^n \sum_{s_i=\pm 1}
\int_0^Ldx_i\int_0^T dt_i \, A(\{x_i,t_i,s_i\}),
\label{expansion}
\end{equation}
where $(x_i,t_i)$ are the space-time coordinates
of a $s_i$ = $\pm 1$ tunneling event (or ``instanton'') at which
$\Psi_{\pm{\bf m}}$ acts, with $\sum_i s_i $ = 0.  The amplitude
$A$ is a product of a {\it one-event} factor
$\prod_i A_1(x_i,s_i)$
and
a {\it two-event} factor
$\prod_{i<j} A_2(x_{ij},t_{ij},s_{ij}) $, with
$x_{ij}$ $\equiv$ $x_i-x_j$, $t_{ij}$ $\equiv$ $t_i-t_j$,
and $s_{ij}$ $\equiv$ $s_is_j$.
Here $A_1$ = $(|t(x_i)|/\hbar ) \exp is_i \bar{\alpha} (x_i)$,
where $\bar{\alpha}(x)$ = $\alpha (x) - Q_0({\bf m})x$,
with $  Q_0({\bf m})$ =
$\langle 0 | \partial_x \varphi_{\bf m} | 0 \rangle$, and
\begin{equation}
A_{2} =
\prod_{\lambda} \left ({ i
d(x_{ij} - v_{\lambda}t_{ij} )
\over \xi {\rm sgn }(v_{\lambda})} \right )^{\eta_{\lambda}({\bf m})s_{ij}},
\label{a2}
\end{equation}
where $\eta_{\lambda}({\bf m})$ =
$|v_{\lambda}|({\bf O}^T{\bf V}^{-1/2} ({\bf mm}^T){\bf V}^{-1/2}
{\bf O} )_{\lambda\lambda}$ $\ge 0 $, with
$\sum_{\lambda} \eta_{\lambda}({\bf m})$
= $2\Delta ({\bf m})$ and
$\sum_{\lambda} \eta_{\lambda}({\bf m}){\rm sgn}(v_{\lambda})$
= $K({\bf m})$;
$ |d(x_{ij} - v_{\lambda} t_{ij} )|$ $\gg$ $\xi $,
where $d(x)$ $\equiv $ $ (L / \pi ) \sin ( \pi x / L) $, is
assumed in  (\ref{a2}).

Perturbations with $K$ = 0
are mass-generating when relevant (the sine-Gordon model is a
familiar example).
In the clean case, the complex phase of the
factor $A_1$ prevents relevance
of $H'$ unless $Q_0$ =
$\partial_x \alpha $.
For relevance, $A_{2}$ requires
the scaling dimension $\Delta $ to be
sufficiently small.  The critical values of $\Delta $
follow from the behavior of (\ref{expansion})
if $L$ and $T$ are rescaled:
$\Delta < 2$ (clean case), $\Delta < \case{3}{2} $ (random case), and
$\Delta < 1$ (local case)\cite{gs,kf}.

When the $K$ = 0
perturbation is relevant, the pair of
modes that ``split off'' in (\ref{reduced}) become a massive
degree of freedom that is removed from the low-energy theory.
The physical significance of the condition $K$ = 0 is that
it implies $[\partial_{x} \varphi_{\bf m}(x),\Psi_{\pm \bf m}(x') ]$ = 0.
When the perturbation is relevant, the phase gradient
$\partial_x \varphi_{\bf m}$
becomes rigidly locked to the value $\partial_x \alpha$
at large lengthscales.  In the random case,
rigidity occurs in finite domains separated by
pinned
``domain walls'' (solitons), and in the local case,
occurs only near $x_0$.
This  rigidity can only develop if
the perturbation commutes with the phase gradient.

The only other perturbations  that could be relevant ($\Delta < 2$)
have
$|K|$ = 2, but are {\it not} mass-generating;
KFP\cite{kfp} argue that in the random case,
a $|K|$ = 2 perturbation with $\Delta  < \case{3}{2}$
renormalizes $\bf V$ so  $\Delta \rightarrow 1$,
its minimum  value.  KFP's arguments seem based in part on
the  familiar renormalization group (RG) treatment of the $K$ = 0 case,
but the  {\it clean} $|K|$ = 2 case\cite{promise} seems
to behave very differently, raising doubts
about such analogies.

If a (primitive)
${\bf m}$ with $K({\bf m})$ = $q({\bf m})$ = 0 exists, the
stability of the Luttinger liquid depends on the non-universal
Hamiltonian coupling matrix ${\bf V}$, which determines
$\Delta ({\bf m})$.    On the other hand, if the
matrix ${\bf K}_0$
{\it does not represent zero}, the edge
is absolutely stable (T-stable), independent of ${\bf V}$.  If
${\bf K}$ is  a definite matrix ($|\sigma (K)|$ = $n$),
the edge is maximally chiral and T-stable.  This  is
sufficient but {\it not} necessary: the condition given
here characterizes $T$-stability in its most general form.

Clearly, if ${\bf K}_0$ is definite, T-stability
is present, as in the case of the $\nu$ = 2/3 edge considered in
\cite{kfp}.
If ${\bf K}_0$
is {\it indefinite} (but  $\sigma^H \ne 0$, so it is nonsingular),
its dimension $n_0$ = $(n-1) \ge 2 $.
If $n_0$ = 2, the system is T-stable if and only if
$-\det K_0$ is not a
perfect square\cite{wat}.  On the other hand, if $n_0$ $>$ 4, an indefinite
${\bf K}_0$ {\it always} represents zero\cite{wat}.
If  $n_0$ is 3 or 4,
an indefinite ${\bf K}_0$ does not represent zero
if it fails to represent zero ``$p$-adically''\cite{wat}
when $p$ is 2, or any
odd prime factor of $\det {K}_0$; this property depends on
the {\it rational equivalence class} of ${\bf K}_0$, and can
be quickly tested by a finite calculation {\it without} searching for
a vector ${\bf m}$ with which ${\bf K}_0$ represents zero.

The formalism presented here allows the
analysis of the edge between Abelian Hall states
$\{{\bf K}^A, {\bf q}^A\}$ and $\{{\bf K}^B,{\bf q}^B\}$.   The first
step is to form the direct sum ${\bf K}$ = ${\bf K}^A \oplus (-{\bf K}^B)$,
${\bf q}$ = ${\bf q}^A \oplus {\bf q}^B$, and reduce the edge theory
to a primitive form.
At an edge between Abelian Hall states with the {\it same}
Hall constant, the Luttinger liquid has $\sigma^H$ = 0,
and ${\bf K}$ can be put into the form
(\ref{reduced}), with ${\bf q}'$ = ${\bf 0}$.
A gap for making charged excitations can open, leaving
a residual neutral gapless
Luttinger liquid.
A RG treatment of the instabilities at
edges between various Abelian Hall states
will be presented
elsewhere\cite{promise}.

The parameters ${\bf V}$
of the Luttinger liquid that determine $\Delta ({\bf m})$
are determined by the internal structure of the edge,
and  can change through non-linear deformations.
It is tempting to speculate that if an edge can exhibit an instability
that reduces the number of edge modes, it will relax so
that the instability occurs, so
only a T-stable edge is robust.   Since T-stable theories are so stable, it
seems natural to conjecture  that only  Abelian Hall
states with this property occur in nature.

In the hierarchy theory,
$q_0 $ = 1, and ${\bf K}_0$ can be  represented\cite{hald,read2} as a
tridiagonal matrix with off-diagonal elements all taking the value 1,
and diagonal elements $(p_1,p_2, \ldots ,p_{n-1})$, where $p_i$ are non-zero
even integers.  At each level of the hierarchy, the most stable
condensates
of quasiparticles correspond to $|p_i|$ = 2 , and are
stabilized by contact repulsion (these may be
called the {\it principal} daughter states).

The $T$-stability condition can now be applied
to determine the allowed {\it signs} that the $p_i$ can have.
First, $\pm (2,2,2,\ldots )$ represents the diagonal elements when
${\bf K}_0$ is definite.  When it is indefinite, $\pm (2,-2)$ is stable;
for $(n-1) > 2$  the test\cite{wat}  shows  that the {\it only}
T-stable case is  $\pm (2,-2,2)$.  In the hierarchy theory, these
correspond to states with $\nu^{-1} $ = $k_0 \mp \case{2}{5} $,
and $k_0 \mp  \case{5}{12}$, with $k_0$ odd.    Thus in the range
$\case{1}{3} \le \nu < \case{1}{2}$, the {\it only} topologically
stable families of spin-polarized (or one-component)
hierarchy states with principal descendants
are the familiar infinite-length sequence  $(\case{1}{3}, \case{2}{5},
\case{3}{7},\case{4}{9},\case{5}{11},\ldots )$, and the
length-4 sequence $(\case{1}{3}, \case{2}{5},\case{5}{13},\case{12}{31})$.
Observation of the finite sequence would test
the idea that the physically-realized states are restricted
to topologically-stable (and usually) principal sequences of hierarchy states.

I would like to acknowledge useful discussions with J. H. Conway,
J. E. Moore, K. Yang, D. Khveshchenko, and J. Gruneberg.
I thank J. Polchinski for pointing out an error in an earlier
version of the formula (\ref{a2}).
This work was supported in part by NSF-DMR-9224077.

\end{document}